# Complex structural dynamics of bismuth under laser-driven compression

Jianbo Hu,[1, 5, †] Kouhei Ichiyanagi,[2] Tomoaki Doki,[1] Arihiro Goto,[1] Takayuki Eda,[1] Katsura Norimatsu,[1] Shinichi Harada,[1] Dai Horiuchi,[1] Yuki Kabasawa,[1] Shingo Hayashi,[1] Shin-ichi Uozumi,[1] Nobuaki Kawai,[3] Shunsuke Nozawa,[4] Tokushi Sato,[4] Shin-ichi Adachi,[4] and Kazutaka G. Nakamura[1, ‡]

[1]*Materials and Structures Laboratory, Tokyo Institute of Technology, R3-10, 4259 Nagatsuta, Yokohama 226-8503, Japan*

[2]*Graduate School of Frontier Sciences, the University of Tokyo, 5-1-5 Kashiwanoha, Kashiwa 277-8562, Japan*

[3]*Institute of Pulsed Power Science, Kumamoto University, 2-39-1 Kurokami, Kumamoto 860-8555, Japan*

[4]*Photon Factory, High Energy Accelerator Research Organization, 1-1 Oho, Tsukuba 305-0801, Japan*

[5]*Laboratory for Shock Wave and Detonation Physics Research, Institute of Fluid Physics, China Academy of Engineering Physics, P. O. Box 919-102, Mianyang 621900, China*

With the aid of nanosecond time-resolved X-ray diffraction techniques, we have explored the complex structural dynamics of bismuth under laser-driven compression. The results demonstrate that shocked bismuth undergoes a series of structural transformations involving four solid structures: the Bi-I, Bi-II, Bi-III and Bi-V phases. The transformation from the Bi-I phase to the Bi-V phase occurs within 4 ns under shock compression at ~11 GPa, showing no transient phases with the available experimental conditions. Successive phase transitions (Bi-V→Bi-III→Bi-II→Bi-I) during the shock release within 30 ns have also been resolved, which were inaccessible using other dynamic techniques.

[†] Present address: Arthur Amos Noyes Laboratory of Chemical Physics, California Institute of Technology, Pasadena, California 91125, USA.
[‡] Author to whom correspondence should be addressed. Electrical mail: nakamura.k.ai@m.titech.ac.jp.

Bismuth (Bi) has one of the most complicated phase diagrams.[1,2] Besides the possible liquid polymorphism[3], Bi has around 10 polytypes as well as anomalous melting. This complexity has severely hindered the understanding of its structural dynamics under compression. On the other hand, it has also stimulated ambitious studies to overcome the difficulties. Both the dynamic high-pressure and static high-pressure communities have devoted effort to this topic.[4-19] Significant progress has been made in the last decade in identifying the crystal structures of the high-pressure phases[17,18] and understanding the transformation kinetics.[13,14] The identified host-guest incommensurate structure of the Bi-III phase is quite unusual, constituting a significant challenge for theoreticians.[20,21] The kinetic process of the Bi-I to Bi-II transition, which has been subjected to extensive studies, has only recently been understood semi-quantitatively through a newly developed ramp compression technique that was used to study the role played by the compression rate and helped to interpret the previous experimental observation.[13,14] However, knowledge of the structural dynamics of Bi is still limited, especially under shock compression. The sequence of shock-induced polymorphous transformations beyond the Bi-I to Bi-II transition has never been systemically identified. Furthermore, the dynamics during shock release has proved almost unreachable owing to the complexity arising from the quasielastic release effect and release-induced multiple phase transitions.[11] In fact, even for the simpler elements, the release dynamics associated with a single phase transition is difficult to resolve using previously available techniques.[22,23] It is also important to study the dynamics of structural phase transitions in bismuth because these transition pressures are used as pressure standards in static high-pressure experiments. Developing experimental techniques to study the structural dynamics during both compression and release is interesting not only for fundamental understanding but also for applications, such as quenching high-pressure phases to ambient pressure.[24]

Time-resolved X-ray diffraction techniques,[25-27] owing to their ability to characterize the crystal structure in real time, may provide the possibility to trace the complex structural dynamics during the entire shock-release process. In this Letter, we have employed a single-shot nanosecond time-resolved X-ray diffraction technique, based on synchrotron radiation, to study the structural dynamics of Bi under laser-driven compression. The results clearly show the sequence of structural transformations during compression and release. Especially, successive phase transitions induced by release have been revealed unambiguously.

The experiments were performed on the beamline NW14A at the Photon Factory Advanced Ring, KEK.[28,29] The sample was a 20 μm thick foil of polycrystalline bismuth that was 99.97 % pure, obtained from Goodfellow Cambridge Limited (UK). The experimental technique has been described in Ref. [30] in detail. A single-shot laser-pump/X-ray probe technique with the time resolution of 2 ns was employed to detect the transient structure of the laser-shocked Bi. The laser pulse used for compression had a pulse energy of 1.0 J and a pulse width (τ) of 8 ns and was focused onto the target assembly with a spot diameter (full width at half maximum) of 0.48 mm. The X-ray probe pulse was normally incident to the target with the wavelength of 0.79 Å and a focal spot of 0.45 × 0.22 mm². As compared to Ref.[30], an energy bandwidth (ΔE/E) of 1.4 % of the X-ray pulse was used to improve the angular resolution at the expense of the photon flux. The target assembly had a plasma-confined scheme, composed of three glued layers: the backup plastic film (25 μm), the Al ablation foil (3 μm) and the Bi foil. The experimental data were collected and analyzed as in Ref. [30]. The peak shock pressure was estimated using a Fabbro-Devaux model[31]

$$P(\text{kbars}) = 0.1 \left(\frac{\alpha}{2\alpha+3}\right)^{1/2} Z^{1/2} \left(\text{g}/\text{cm}^2\text{s}\right) I_0^{1/2} \left(\text{GW}/\text{cm}^2\right) \quad (1)$$

where $Z = 2Z_1Z_2/(Z_1 + Z_2)$ and $Z_i$ is the shock impedance of the ablation and confining materials. In this case, $Z_{Al} = 1.5 \times 10^6$ g/cm²s and $Z_{Polymer} = 0.2 \times 10^6$ g/cm²s as estimated from Hugoniot data.[32] $I_0$ is the laser energy, $\alpha = 0.2$ the corrective factor of water confinement, which is used to approximate polymer confinement, as the impedance of the polymer is close to that of water. The estimated shock pressure is ~11 GPa, higher than the reported phase transition pressure of 7.7 GPa for the Bi-V phase.[17] For each delay between the pump and probe pulses, a fresh target assembly was used because of the penetration induced by the laser shock.

Debye-Scherrer diffraction patterns before and after laser pumping were recorded for each delay time, Δt. The former was used to check the consistency of the target assemblies and act as the reference for the latter. As an example, Figure 1(a) and 1(c) show the diffraction patterns from the pristine sample and the shocked sample at Δt=12 ns. Changes are evident in the ring patterns, not only through changes in the brightness of the diffraction rings. New rings have also appeared, indicating the possible emergence of a new phase after the laser shock. The normalized diffraction intensity profiles obtained by azimuthally averaging of the Debye-Scherrer diffraction patterns are shown in Fig. 1(b) and 1(d), providing more direct information.[30] The diffraction peaks before the laser pump (Fig. 1(b)) correspond to the Bi-I phase (R-3m), while the new peaks in Fig. 1(d) are almost consistent with the Bi-III phase (host-guest structure). This means that the laser-shocked Bi has undergone a phase transformation during either compression or release.

To systematically map the structural dynamics after the incident laser pulse, a series of pump-probe experiments at different delay times were carried out from -2 ns to 36 ns. The temporal evolution of the normalized diffraction intensity profile is shown in Fig. 2(a). The selected signals at Δt=0 ns, 4 ns, 14 ns, 20 ns and 30 ns are shown in Fig. 2(b), and are compared with the calculated diffraction peaks of Bi-I to Bi-V phases. The crystal parameters for the calculation are taken from Ref. [33], in which the compressibility of the Bi-I and Bi-III phases has been considered while that of the Bi-II phase has been ignored owing to its narrow pressure range. Here, the zero delay is defined by a non-observable change in either intensity or diffraction angle of the diffraction intensity profiles with and without laser irradiation. The comparison demonstrates that a series of phase transitions have been observed after laser irradiation. The observed new peaks correspond to the Bi-V phase (Im-3m) at Δt=4 ns, to the Bi-III phase at Δt=14 ns, and to the Bi-II phase at Δt=20 ns. At Δt=30 ns, the shocked sample transforms back to the Bi-I phase. Note that to identify the new structures, not only the peak positions but also the change in peak intensities have been taken into account. Owing to the closeness of the diffraction peaks, it is not a trivial task to distinguish the Bi-II and Bi-III phases in Fig. 2(b). We therefore analyzed the diffraction intensity profiles using curve fitting with a superposition of Bi-I and Bi-II or Bi-III phases. Gaussian functions with a peak width of 0.7 degrees were used as the fitting function for the Bi-II and Bi-III phases and 0.4 degrees for the Bi-I phase. Figure 3(a) shows the diffraction intensity profile obtained at Δt=14 ns and the best fit from the Bi-I and Bi-III phases. In this calculation, the intensity ratio of the Bi-III to Bi-I phases was 0.83. Figure 3(b) shows the diffraction intensity profile obtained at Δt=20 ns and its best fit from the Bi-I and Bi-II phases with an intensity ratio (Bi-II/Bi-I) of 0.37. Both the diffraction intensity profiles at Δt=14 ns and Δt=20 ns are well reproduced from the superposition of Bi-I and Bi-III and Bi-I and Bi-II phases, respectively. We furthermore calculated the convolution of the diffraction intensity profiles before and after pumping. This procedure can reduce the effect of the noise and thus provide a better comparison between the experimental and fitted results. As seen in the insets of Fig. 3, it clearly shows that the Bi-III phase dominates the variation of the diffraction intensity profile at Δt=14 ns and the Bi-II phase does at Δt=20 ns. To summarize, the laser-shocked Bi underwent the phase transitions from Bi-I to Bi-V during compression, and then relaxed to Bi-III, to Bi-II, and finally back to Bi-I during release.

We extracted the structural information at each delay time by comparing the experimental diffraction profile with the calculated diffraction peaks. The structural information at different delay times is schematically summarized in a time-pressure plot, Fig. 4. The coexistence of phases occurs at some delay times and so only the higher-pressure phase is listed. The possible pressure hysteresis of phase transitions during release has also been ignored. It is well known that the output pressure pulse in a plasma-confined geometry is determined by the input laser pulse and the properties of relevant materials. In the Fabbro-Devaux model, a step-like laser pulse is assumed to simplify the solution.[31] For a Gaussian laser pulse employed in the experiments, we may approximate the pressure profile by the convolution of the Fabbro-Devaux model (Eq.1) and a Gaussian function in the duration of the laser pulse ($t \leq \tau$). At times longer than the pulse duration ($t > \tau$), the plasma undergoes the uniaxial relaxation until the edge rarefaction arrives at the center. The pressure profile can be approximated by[34]

$$P = P_S \left(\frac{\tau}{t}\right)^{2/3} \qquad (2)$$

where $P_S$ is the pressure at the end of the pulse duration. The calculated shock pressure profile is also shown in Fig. 4. It is not difficult to see that in Fig. 4 the time evolution of the crystal structure of laser-shocked Bi follows the variation of the shock pressure in the Bi sample. The Bi-I phase is transformed into the Bi-V phase during compression and then sequentially into Bi-III, Bi-II and Bi-I phases during release.

A direct transition from Bi-I to Bi-V is, to some extent, unexpected because no direct transformation path has been proposed so far.[35] There are two possibilities. One is simply that the time the sample spends as the Bi-II and Bi-III phases during compression is too short to be resolved with currently available experimental conditions. The other arises from the reconstructive character of the Bi-I to Bi-II transition, which requires an incubation time of some tens of nanoseconds.[13,14] Thus the Bi-I phase would be overshocked to the Bi-V phase, although the displacive transformation path between the two phases is still unknown. To distinguish between two possibilities it would be necessary to conduct time-resolved X-ray diffraction experiments at the shock pressure lower than the transition pressure of the Bi-V phase.

As mentioned above, the structural dynamics of Bi during the shock release was previously too complex to analyze.[11] Quantitative analysis is hindered not only by the limited signal-to-noise ratio, but also by the complexity of the diffraction pattern induced by coexisting phases and pressure inhomogeneity. Our experimental results, however, clearly showed successive phase transitions from Bi-V to Bi-III to Bi-II to Bi-I during release. The transitions from Bi-V to Bi-III to Bi-II, due to the displacive mechanism,[35] are extremely rapid as they are associated with a considerable pressure change (corresponding to the angle shift in the diffraction peaks). The Bi-II to Bi-I transition, although being present over a narrow pressure range, takes a relatively long time. This is because the reconstructive mechanism of the transition requires the nucleation and growth of the new phase and is time-consuming.[13,14] The absence of Bi-IV implies that the Bi-IV phase might be a low-temperature phase, consistent with the results of McMahon *et al.*.[17] This study demonstrates the power of time-resolved X-ray diffraction techniques to study complex structural dynamics during either shock compression or release. Actually, the shock release is essentially a ramp process, and as such avoids the technical challenge of generating high-quality ramp waves.

In summary, we have systemically studied the structural dynamics of Bi under laser-driven compression using a nanosecond time-resolved X-ray diffraction technique. A series of structural transformations have been observed during both compression and release. This indicates the ability of time-resolved X-ray diffraction technique to study complex structural dynamics of materials under dynamic compression, which can bridge the gap between dynamic and static high-pressure communities.

# List of figure captions

**FIG. 1** (color online) Debye-Scherrer diffraction patterns from (a) the pristine sample and (c) the shocked sample at $\Delta t$=12 ns. (b) and (d) show the integrated intensity profiles from the pattern (a) and (c), respectively, normalized to the total intensity. The two series of bars in (b) and (d) respectively show the calculated diffraction peaks of the Bi-I (green) and Bi-III (blue) phases.

**FIG. 2** (color online) (a) Time evolution of the normalized diffraction intensity profiles for delay times from -2 ns to 36 ns. (b) Selected signals at the delay times of 0 ns, 4 ns, 14 ns, 20 ns, and 30 ns for comparison with the calculated diffraction peaks. The series of bar show the calculated diffraction peaks of the Bi-I (green), Bi-II (grey), Bi-III (blue), and Bi-V (azure) phases. The wide bars of Bi-III are due to the compressibility of the Bi-III phase.

**FIG. 3** (color online) (a) Diffraction intensity profile after laser pumping (red dots) at $\Delta t$=14 ns and the fitted curve (blue line) from the superposition of Bi-I and Bi-III phases. The profiles of Bi-I and Bi-III phases are shown as green and grey dashed lines, respectively. The sticks represent the diffraction peaks of the Bi-III phase. (b) Diffraction intensity profiles after laser pumping (red dots) at $\Delta t$=20 ns and the fitted curve (blue line) from the superposition of Bi-I and Bi-II phases. The profiles of Bi-I and Bi-II phases are shown as green and grey dashed lines, respectively. The sticks represent the diffraction peaks of the Bi-II phase. The insets show the experimental and fitted results of the convolution of the diffraction profiles before and after pumping.

**FIG. 4** (color online) Schematic of the time evolution of the crystal structure of laser-shocked Bi. The shaded areas with different color indicate the presence of each phase. The dashed line represents the calculated output pressure pulse driven by the laser-generated plasma. The guest Bi atoms in the Bi-III phase are designated by the orange balls.

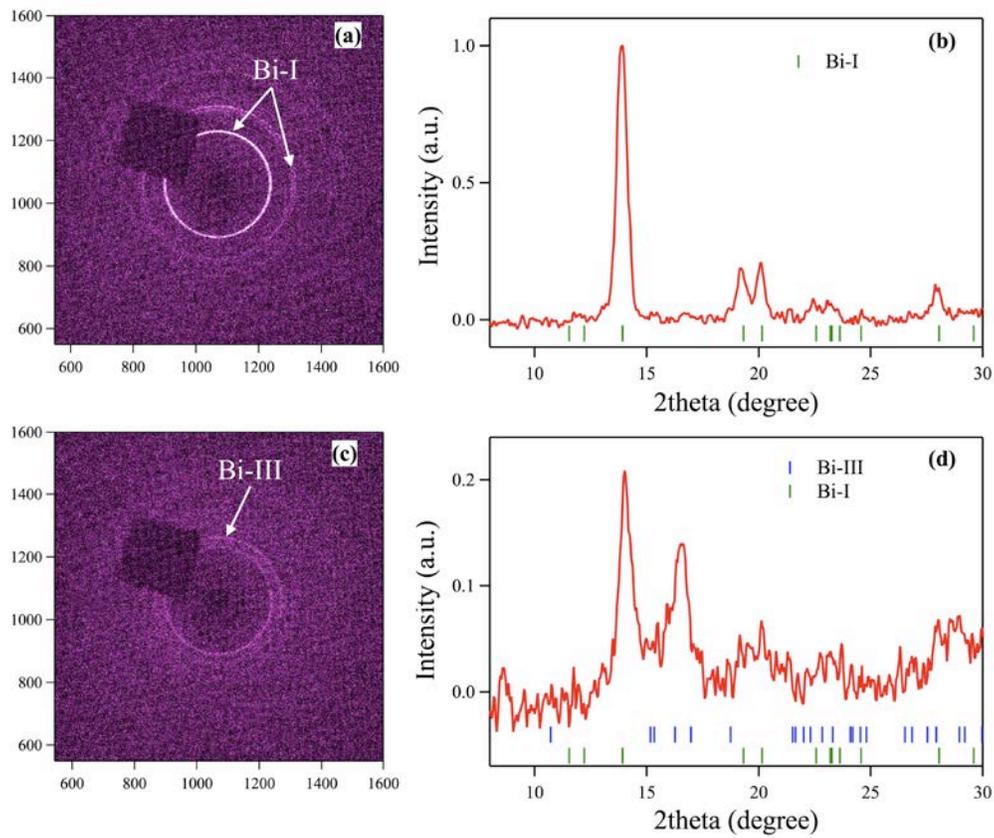

**FIG. 1 Hu** *et al*.

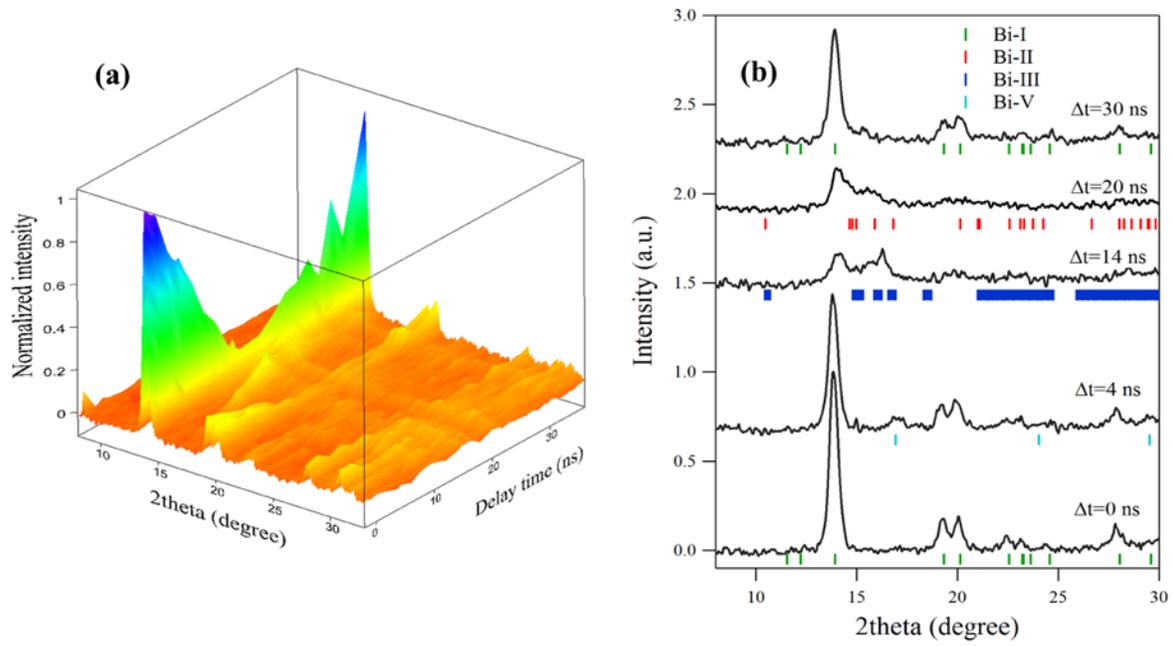

**FIG. 2  Hu *et al*.**

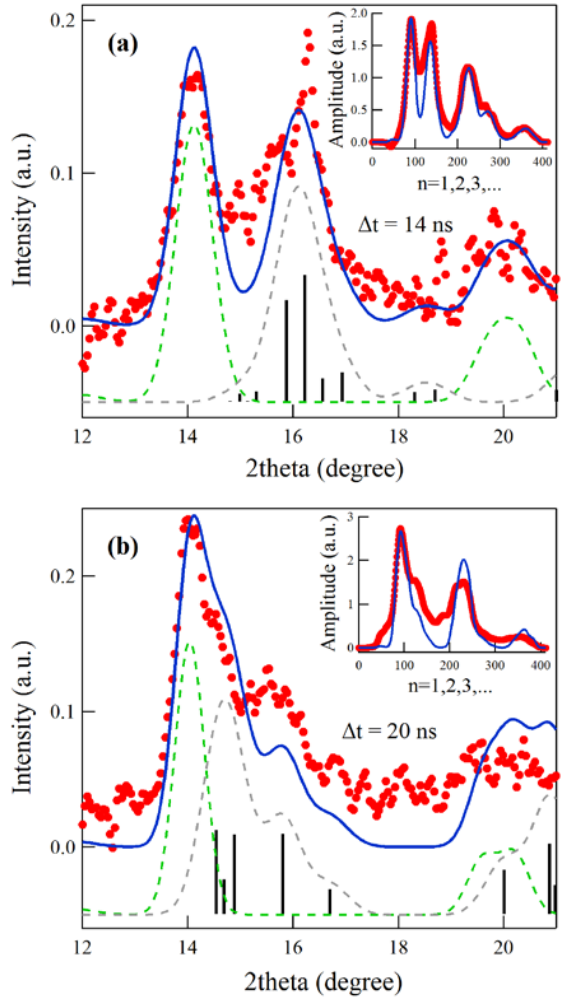

**FIG. 3 Hu *et al*.**

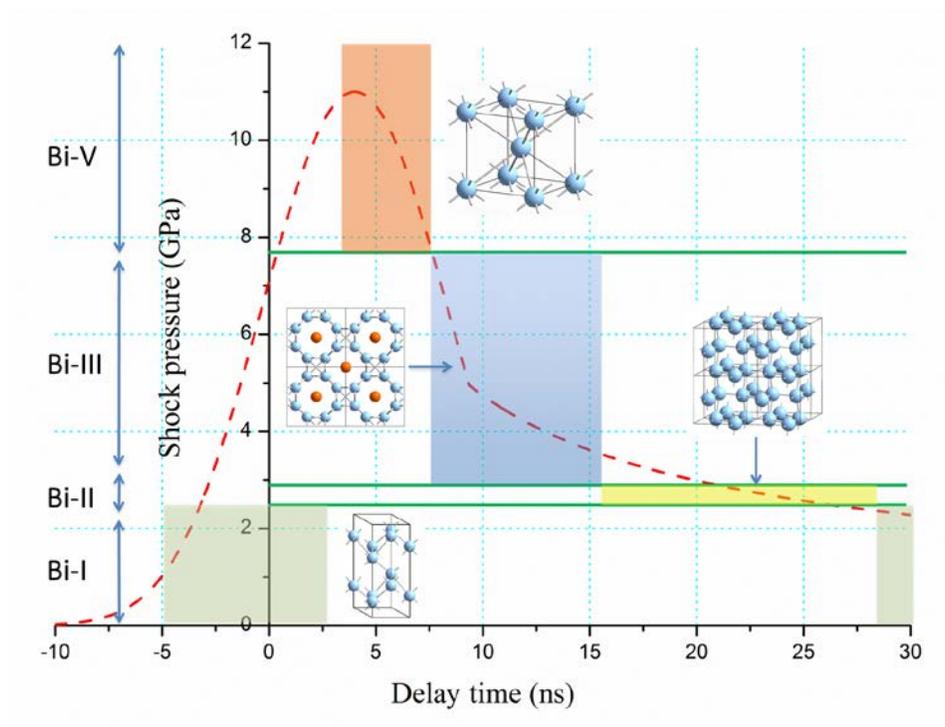

**FIG. 4 Hu** *et al.*